\documentclass[aps,prd,superscriptaddress,nofootinbib]{revtex4}
\usepackage{amsmath, amsthm, amsfonts}
\usepackage{graphicx}
\usepackage{natbib}
\begin{document}
\title{Top-Hat Spherical Collapse with Clustering Dark Energy. I. Radius Evolution and Critical Contrast Density}
\author{D. Herrera} 
\email{duvanh@if.ufrj.br} 
\author{I. Waga} 
\email{ioav@if.ufrj.br} 
\author{S.E. Jor\'as} 
\email{joras@if.ufrj.br}
\address{Instituto de F\'\i sica, Universidade
Federal do Rio de Janeiro\\{C. P. 68528, CEP 21941-972, Rio de Janeiro, RJ,
Brazil}}

\begin{abstract}
Understanding the influence of dark energy on the formation of structures is  currently a major challenge in Cosmology, since it can distinguish otherwise degenerated viable models. In this work we consider the Top-Hat Spherical-Collapse (SC) model with dark energy, which can partially (or totally) cluster, according to a free parameter $\gamma$. The {\it lack of} energy conservation has to be taken into account accordingly, as we will show. 
We determine characteristic quantities for the SC model, such as the critical contrast density and radius evolution, with particular emphasis on their dependence on the clustering parameter $\gamma$.

\end{abstract}

\maketitle

\section{Introduction}

Recent results \cite{ref19,p18} from independent cosmological observations --- such as anisotropies in the Cosmic Microwave Background (CMB), Baryon Acoustic Oscillations (BAO),  type-Ia Supernovae (SNe Ia) and the Large-Scale Structure of the Universe (LSS) --- imply that the Universe is speeding up. \cite{ref4,ref5}. 
The responsible for this effect  is dubbed ``dark energy'' (DE), whose physical nature is still unknown. If we model dark energy as a fluid, according to General Relativity, it needs to have negative pressure. In particular, the cosmological model that better fits observations is the cold-dark-matter with Cosmological-Constant model ($\Lambda$CDM). 
However, this model presents difficulties at theoretical level \cite{cc,ref9}, motivating the search for alternatives such as quintessence \cite{qui,Wetterich,Frieman,qui2}, phantom dark energy \cite{fantasma}, k-essence \cite{ke}, decaying vacuum models \cite{v2,v1} or even modifications of General Relativity, such as $f(R)$ theories \cite{teorias}, among others. A great difficulty is that many of these models behave very similarly to $\Lambda$CDM  at the background level, making it difficult to distinguish them through cosmological kinematical tests (those that depend essentially only on distance). Therefore, it is crucially important to study the evolution of perturbations and the structure formation in those models, where they are expected to have different (and measurable) consequences from those obtained by  $\Lambda$CDM.

The simplest way to study the structure formation with dark energy is through the  Top-Hat Spherical-Collapse (SC) approach, which was initially used in Einstein-de Sitter (EdS) background (as an useful benchmark since it yields an exact analytical result for the critical density), in the standard cold-dark-matter scenario \cite{ref31}, and later in $\Lambda$CDM \cite{c1}. The SC model has also been extended to quintessence fields \cite{c3,Creminelli:2009mu,qc2}, decaying vacuum models \cite{c4},  $f(R)$ theories \cite{ref36,f1,koop1,yo,campoc}, DE with constant equation-of-state (EoS) models  \cite{Abramo:2007iu,2010JCAP...10..028L,c6,c5}, coupled DE models \cite{Sapa,Valeria}, and  agegraphic DE cosmologies \cite{Rezaei}.   
In particular, Ref.~\cite{Abramo:2007iu} investigated constant phantom, constant non-phantom and varying DE EoS parameter, always assuming that the latter is the same both inside and outside the collapsed region\footnote{As we will show further down, this assumption is equivalent to requiring that the DE EoS parameter is equal to its speed of sound squared: $w=c_s^2$.}. These authors have focused only in the limiting cases, namely,  fully clustered and completely homogeneous DE.

In Ref.~\cite{c5} the SC model with fully clustered DE is considered assuming a linear relation between the matter contrast density and the DE one, according to a free parameter $r$. In Ref.~\cite{c5}, as well as in \cite{2010JCAP...10..028L}, it is also assumed that the DE EoS is the same inside and outside the collapsed region. 

In this work we relax the aforementioned hypotheses and generalize some of those results. Following the {\it Ansatz} suggested in Ref.~\cite{vira} (see also \cite{c3}),  we investigate the SC model with DE, assuming that it can cluster partially or totally, according to a normalized parameter: if $\gamma=0$, DE is fully clustered; if $\gamma=1$, DE is completely homogeneous. 
This paper is organized as follows. In Section~\ref{eqs}  we show the basic equations that describe the SC model with dark-energy perturbations.
We apply the so-called differential-radius method, which has been shown \cite{yo} (see also \cite{Pace}) to be more robust than the constant-infinity method --- which uses a fixed large value for the local overdensity as a threshold for indicating a collapsed structure. The former method, on the other hand, follows the difference between the background scale factor and the collapsing bubble radius (also known as local scale factor). 

In Subsection~\ref{sec_radius} we analyze the radius evolution of the collapsing spherical region. In Section~\ref{sec_critical} we determine the critical contrast density. We conclude in Section~\ref{sec_conclusions}.

\section{Spherical collapse with dark energy perturbations}
\label{eqs}
For a flat, homogeneous and isotropic universe with dark matter and dark energy, the Einstein equations are given by:
\begin{eqnarray}
\label{1}
 \left(\frac{\dot{a}}{a}\right)^{2} &\equiv& H^{2} = \frac{8\pi G}{3}\left(\bar{\rho}_{m}+ \bar{ \rho}_{de}\right), \\
%
\label{2}
 \frac{\ddot{a}}{a} &=& -\frac{4\pi G}{3}\left[\bar{\rho}_{m}+ (1+ 3 w) \bar{\rho}_{de}\right]. 
\end{eqnarray}
In the equations above, $a$ is the scale factor, $H$ is the Hubble parameter, $w\equiv \bar p_{de}/\bar \rho_{de}$ is the EoS parameter of DE (assumed to be constant), and $\bar{\rho}_{m}$, $\bar{\rho}_{de}$ and  $\bar p_{de}$ are the (background) energy densities of matter and DE and the DE pressure, respectively. A dot over a given quantity denotes its time derivative.

Assuming that both dark matter and DE interact only gravitationally and are separately conserved, we get
\begin{eqnarray}
\label{3}
&& \dot{\bar{\rho}}_{m} +3H\bar{\rho}_{m}=0, \\
%
\label{3.1}
&& \dot{\bar{\rho}}_{de} +3H(1+w)\bar{\rho}_{de}=0. 
\end{eqnarray}

Here we investigate the nonlinear evolution of the gravitational collapse and, to this aim, we consider the Top-Hat Spherical-Collapse (SC) model. The SC model considers a spherical region with a top-hat profile and  uniform density  $\rho(t) = \bar{\rho}(t) + \delta \rho(t)$, immersed in a homogeneous universe with energy density $ \bar{\rho}(t)$. Here $\delta \rho$ initially is a small perturbation of the background fluid energy density. We suppose that this region also contains nonrelativistic matter ($p_m=\bar p_m=0$)  and DE. Such a spherical region can be described as a separated universe with (local) scale factor $r$. The acceleration equation for this region is given by:
\begin{equation}\label{4}
\frac{\ddot{r}}{r}=-\frac{4\pi G}{3}\left(\rho_{m}+ \rho_{de}+3 p_{de}\right),
\end{equation}
where $p_{de} (t)= \bar p_{de}(t)+  \delta p(t)$ is the DE pressure inside the spherical region and $ \delta p(t)$ a small pressure perturbation. The DE EoS parameter in the spherical region is given by \cite{Abramo:2007mv}
\begin{equation}\label{4a}
w^{c} \equiv \frac{p_{de}}{\rho_{de}} = w+\frac{(c_s^2 -w)\delta_{de}}{1+\delta_{de}},
\end{equation}
where the superscript ``$^c$'' stands for ``clustered'', $c_s^2\equiv \delta p_{de}/\delta \rho_{de}$ is the DE sound speed squared (assumed to be constant) and $\delta_{de}$ is the DE density contrast (see its definition below). Note that only if $c_s^2=w$ (or homogeneous DE, i.e., $\delta_{de}=0$) the DE EoS parameter in the collapsing region is equal to that of the background ($w^{c}= w$). 

Due to its standard attractive character, dark matter always tends to cluster, so the local continuity equation takes a similar form as the continuity equation for the background fluid, that is:
\begin{equation}\label{6}
\dot{\rho}_{m}+ 3\frac{\dot{r}}{r}\rho_{m}=0,
\end{equation}
where $r$ is the local scale factor. Of course, it is clear that dark matter will actually cluster only if the initial $\delta \rho_{m}$ is large enough to overcome the effects from both the background expansion and DE. In the present work we assume that DE can also collapse --- although not necessarily together with the matter content, since it can flow away from the collapsing sphere. This is precisely the reason for the lack of energy conservation in the perturbed region. Therefore, we parameterize such physical phenomenon writing the local continuity equation for DE as \cite{vira} (see also \cite{c3}) :
\begin{equation}
\label{7}
 \dot{\rho}_{de}+ 3(1+w^{c})\frac{\dot{r}}{r}\rho_{de}=\gamma \Gamma\, ,\quad \textup{with} \quad   0 \leq  \gamma \leq 1 \, , \\
\end{equation}
where
\begin{equation}
\label{8}
 \Gamma\equiv 3(1+w^{c}) \left( \frac{\dot{r}}{r} -\frac{\dot{a}}{a} \right) \rho_{de}\, .
\end{equation}
Here, $\Gamma$ describes the leaking of DE away from the spherical collapsing region and  $0\leq\gamma\leq1$ is the aforementioned clustering parameter. 
The non-clustering, i.e, homogeneous DE corresponds to $ \gamma = 1 $. Notice that in this case, we have $\rho_{de} \propto \exp[-3\int (1+w^c) da/a]$ while $ \bar{\rho}_{de}$ scales as $ \bar{\rho}_{de} \propto a^{-3(1+w)}$. So, in principle, even if the DE energy densities were initially equal, they would evolve differently. However, as we will show further down, when $\gamma =1$, in the linear regime, there is no growing mode and  $\delta \rho_{de}$ rapidly tends to zero. Therefore, it is not possible to distinguish in this case the behavior of the DE inside and outside the spherical region:  $\rho_{de} = \bar{\rho}_{de}$ and consequently $w^{c}= w$. In this case (and also for $\gamma>0$) the total energy of the system is not conserved \cite{vira}. In contrast, the case of full clustering, i.e, when $\gamma=0$, ensures that  $\rho_{de}\neq \bar{\rho}_{de}$, such that the spherical region is completely segregated from the background and it is considered an isolated system, which conserves energy. We shall also consider intermediate values of $\gamma$ in our analysis. Notice that, differently from Ref.~\cite{vira}, we are not assuming that the DE EoS is the same inside and outside the collapsing spherical region. As remarked above, this is only the case when dark energy is homogeneous ($\gamma = 1$) or $c_s^2=w$. 

Differentiating twice the density contrast $ \delta_{j}  \equiv \rho_{j} / \bar{\rho}_{j} - 1 $ for both dark matter ($\delta_m$)  and dark energy ($\delta_{de}$) and using the equations above we obtain the following nonlinear evolution equations :
\begin{equation}
\label{9}
 \ddot{\delta}_{m} +2 H\dot{\delta}_{m} - \frac{4  \dot{\delta}^{2}_{m}}{3(1+\delta_{m})}  =\frac{3 H^{2}}{2}(1+\delta_{m})\left(\Omega_{m}\delta_{m} + (1-\Omega_{m})\delta_{de}(1+3 c_s^2)\right),
\end{equation}

\begin{eqnarray}
\label{11a}
\ddot{\delta}_{de} &=&-3 (h-H) (1+w) (1-\gamma) \dot{\delta} _{de}-3 (h (1-\gamma )+
\gamma H) \delta w \dot{\delta} _{de}-
\nonumber \\
&-& 
3 (1+w) (1-\gamma) \left(\dot{h}-\dot{H} \right)
\left(1+\delta _{de} \right)-
\nonumber  \\
&-& 
3 \delta w 
 (1-\gamma) 
\left( \dot{h}(1-\gamma)+\gamma  \dot{H} \right) \left(1+\delta _{de} \right) -
\nonumber \\
&-& 
3 \left(h (1-\gamma)+\gamma
H\right) \dot{\delta} w \left(1+\delta _{de}\right).
\end{eqnarray}

In the expression above, $\delta w \equiv w^{c} - w$ (see eq.~(\ref{4a})), 
\begin{equation}
h \equiv \frac{\dot{r}}{r} =\frac{\dot{\delta} _{de}+3 H \left((1+w) (-1+\gamma \
)+\left(-1-w+\gamma +\gamma  c_s^2\right) \delta _{de}\right)}{3 \
(-1+\gamma ) \left(1+w+\left(1+c_s^2\right)
\delta _{de}\right)},
 \label{12}
\end{equation}
\begin{equation}
\dot{h}=\frac{\ddot{r}}{r}-h^2,
\end{equation}
\begin{equation}
\frac{\ddot{r}}{r}=-\frac{H^2}{2}\bigg[\Omega_m (1+\delta_m)+(1-\Omega_m)\big((1+3c_s^2)\delta_{de}+1+3w\big)\bigg]
\end{equation}
and
\begin{equation}
\label{13}
\dot{H}=-\frac{3}{2} H^2 \left(1+w \left(1-\Omega _m\right)\right).
\end{equation}
Here $\Omega_{m}=\Omega_{m}(t)$ is the background nonrelativistic matter energy-density parameter at the instant $t$. 

In the expressions above we assume, obviously, that $\gamma \neq 1$, since, as mentioned before, if $\gamma= 1$ DE does not cluster. We note that in the particular case in which $c_s^2=w$ \cite{Abramo:2007iu,2010JCAP...10..028L,c5}, such that $\delta w=0$,  eq.~(\ref{11a}) reduces to 
\begin{eqnarray}
 \ddot{\delta}_{de} &+& 2 H\dot{\delta}_{de} -\frac{4+3w-3\gamma(1+w)}{3(1+w)(1-\gamma)} \frac{\dot{\delta}^{2}_{de}}{(1+\delta_{de})}  = \nonumber \\
&=&  \frac{3 H^{2}}{2}(1+\delta_{de})(1-\gamma)(1+w)[\Omega_{m}\delta_{m} + (1-\Omega_{m})\delta_{de}(1+3 w)].
\end{eqnarray}
If we further impose $\gamma =0$, we then recover  Eq. (7) of Ref.~\cite{Abramo:2007iu}  for the case in which $w$ is constant.

\begin{figure}[h]
\centering
\includegraphics[width=\textwidth]{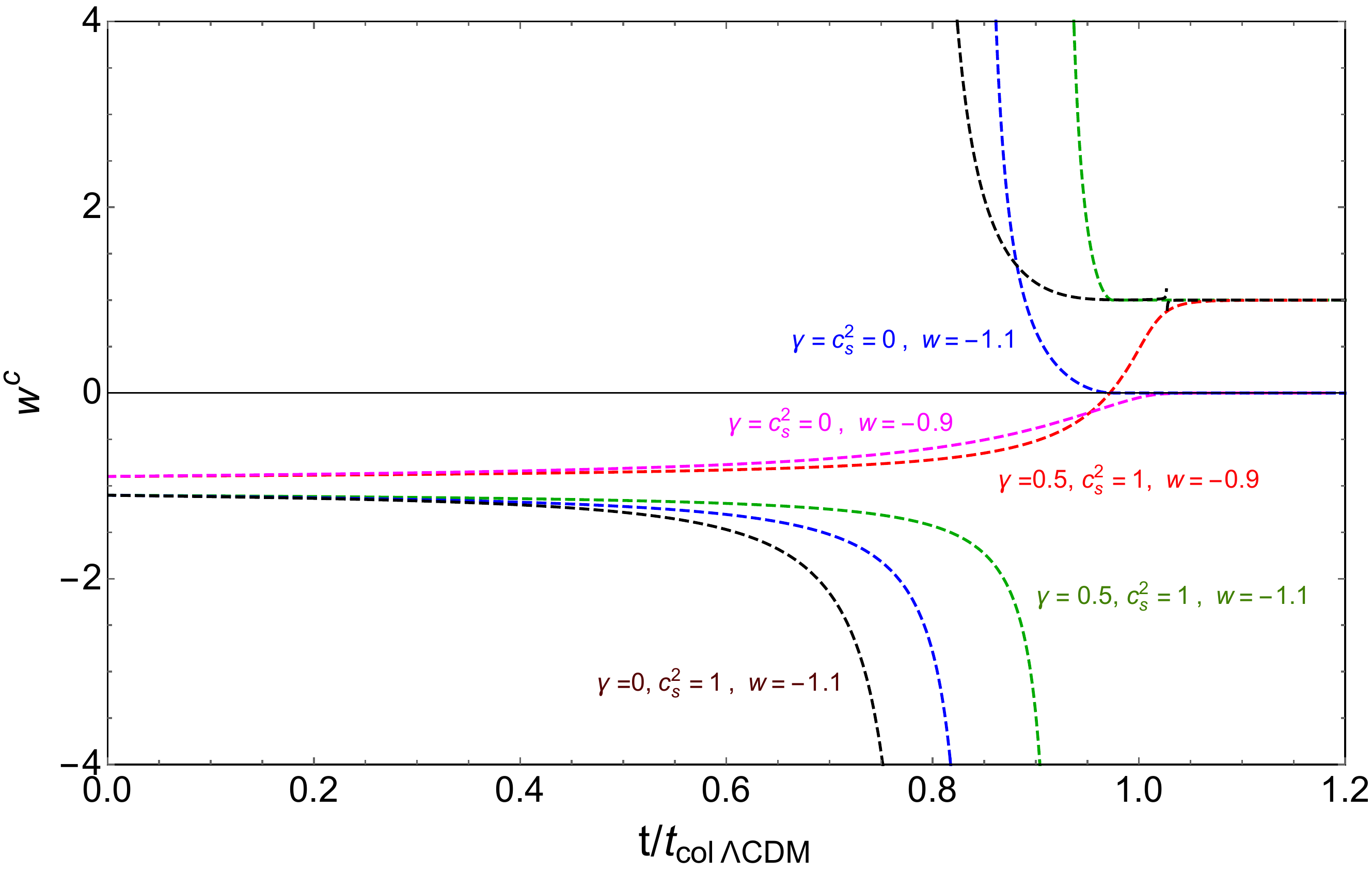}
\caption{Behavior of the equation-of-state parameter inside the bubble ($w^c$) with respect to time, for the labeled parameters.}
\label{wctcol}
\end{figure}

To determine the initial conditions for $\delta_ {m} $ and $\delta_{de}$, we consider the linear approximation of Eqs.~(\ref{9}) and (\ref{11a}) in a matter-dominated universe ($\Omega_m \sim 1$ and $\Omega_{de}\sim 0$):
%
\begin{eqnarray}
\label{16a}
\delta''_{m} &+& \frac{3}{2}\frac{\delta'_{m}}{a}-\frac{3}{2 a^2}\delta_{m}=0 \\
%
\label{17a}
\delta''_{de} &+& \left(\frac{3}{2}- 3(w-c_s^2)\right)\frac{\delta'_{de}}{a}- \nonumber \\
 &-& \frac{3}{2 a^{2}}\left((1+w)(1-\gamma)\delta_{m}+(w-c_s^2)\delta_{de})\right)=0,
\end{eqnarray}
where $'\equiv d/da$ . Since we are interested in the formation of structures, the decreasing mode of the above equations will not be considered. The growing mode solutions are: 
%
\begin{eqnarray}
\label{13a}
\delta_{m}(a) &=& C \, a \qquad \qquad \qquad
{\rm and}\\
\label{14a}
\delta_{de}(a) &=& \frac{(1+w)(1-\gamma)}{1-3(w-c_s^2)}\delta_{m}(a).
\end{eqnarray}
As remarked above, if $\gamma=1$ we obtain $\delta_{de}=0$. We assume in our analysis that $1-3(w-c_s^2)>0$ which implies that for phantom models ($w<-1$) $\delta_{de}<0$ (i.e, there is less dark energy inside the bubble than in the background). 
Note that if $\delta_{de}<-1$, then $\rho_{de}<0$. Although such case is exotic, in principle, it is allowed in some modified gravity models \cite{Nozari_2009}. Whenever $\delta_{de}$ crosses $-1$, which happens only if $w<-1$, then $w^c$ goes from $-\infty$ to $+\infty$  (see  Eq.~(\ref{4a}) and Fig.~\ref{wctcol}). Note, however, that such divergence does not affect the evolution of the bubble, since $w^c$ does not appear explicitly in the equations of motion for the radius $r$ (or, actually, for the variable $y$), as we will show next.

Given the contrast density  for each fluid, the evolution of the {\it local} scale factor is given by Eq.~(\ref{4}), which in terms of  $y \equiv \frac{r}{r_{i}}-\frac{a}{a_{i}}$ can be written as:
\begin{eqnarray}
\label{15a}
 y'' &+& \left(y'+\frac{1}{a_{i}}\right) \left(\frac{H'}{H}+\frac{1}{a}\right) = \\ \nonumber
& & =   -\frac12\left(\frac{H_{0}}{ H a}\right)^{2}\left(y+ \frac{a}{a_{i}}\right) \times \\ \nonumber
& & \quad \times \bigg[\Omega_{m0}a^{-3}\Big(1+\delta_{m}\Big) + \nonumber\\
 && \quad  + \Big(1+3w+(1+3c_s^2)\delta_{de}\Big)\Big(1-\Omega_{m0}\Big)a^{-3(1+w)}\Big(1+\delta_{de}\Big)\bigg] , \nonumber 
\end{eqnarray}
where $\Omega_{m0}$ is the present value of the matter density parameter. An initial condition for Eq.~(\ref {15a}) is naturally $y (a_{i}) = 0 $. To obtain $y '(a_{i})$, we consider that, {\it initially}, the mass of the spherical region is given only by the contribution from dark matter:
 \begin{equation}
 M_i=\frac43 \pi R_i^3 (1+ \delta_{mi})\bar\rho_{mi}.
 \end{equation}
The (possible) contribution from DE is negligible since $\rho_{de}\ll \rho_m$ when the initial conditions are set, in a matter-dominated universe. 

 In the above equations, $R_i \equiv r(t_i) X$ is the physical radius of the collapsing sphere at instant $t_i$,  X is its coordinate radius and $\delta_{mi}$ is the initial matter density contrast.
Since dark matter always collapses (depending, of course, on the initial conditions of the matter perturbations), the mass $M$ inside the spherical region will always be a constant. Thus, we have $y'(a_{i})= -\delta'_{mi}/ [3(1+\delta_{mi})]$. We adopt in our numerical calculations $a_i=10^{-5}$.

\section{Bubble Evolution}

In this section we investigate the bubble evolution, namely its radius as a function of time, and one of the main results from the SC model: the critical density contrast --- a crucial quantity to determine the number of collapsed objects. Throughout the paper we assume that $\Omega_{m0}=0.3$. We also keep the same initial conditions for dark-matter perturbations, such that the collapse in $\Lambda$CDM always occurs at the present time. 

We pay special attention to the dependence of the outcomes in the free parameters of our model: $\gamma$, $c_s^2$ and $w$. Some  situations are particularly interesting and express the richness of the present parametrization:  

\begin{itemize}
    \item $c^{2}_{s}=0$, in which there is no DE pressure perturbation. It is interesting to point out that, in this case, in the final stages of the collapse ($\delta_{de} \rightarrow \infty$), the local dark energy does behave as dark matter, since $w^c \rightarrow 0$ --- see eq.~(\ref{4a}). Note also, from eq.~(\ref{14a}), that for phantom dark energy one will always get $\delta_{de}<0$: there is less dark energy inside the bubble than in the background.
    \item $c^{2}_{s}=w$, which indicates that the clustered DE EoS parameter ($w^c$) and the background one ($w$) are equal.
     \item $c_s^2=\gamma$.  We intend to model a continuous ``turning on'' of the clustering in scalar field models \cite{vira}. In quintessence and k-essence models, usually, two choices are made:
     \begin{itemize}
     \item[a)] $c_s^2=1$, in which case the standard quintessence scalar field (i.e, a minimally coupled scalar field with a canonical kinetic term) does not cluster, remaining homogeneous on subhorizon scales \cite{qui2}, and 
      \item[b)] $c_s^2=0$ (or more generally sub-luminal behaviour) are considered in k-essence scalar fields  \cite{Creminelli:2009mu}.
     \end{itemize}
\end{itemize}

The new parameter $\gamma$ models the lack of energy conservation, which happens whenever a fraction of DE does not cluster.  Note that when $\gamma=1$, results from different $c_s^2$ should coincide, since the latter does not play a role if DE is homogeneous.
    
\subsection{Radius}
\label{sec_radius}

\begin{figure}
\centering
\includegraphics[width=\textwidth]{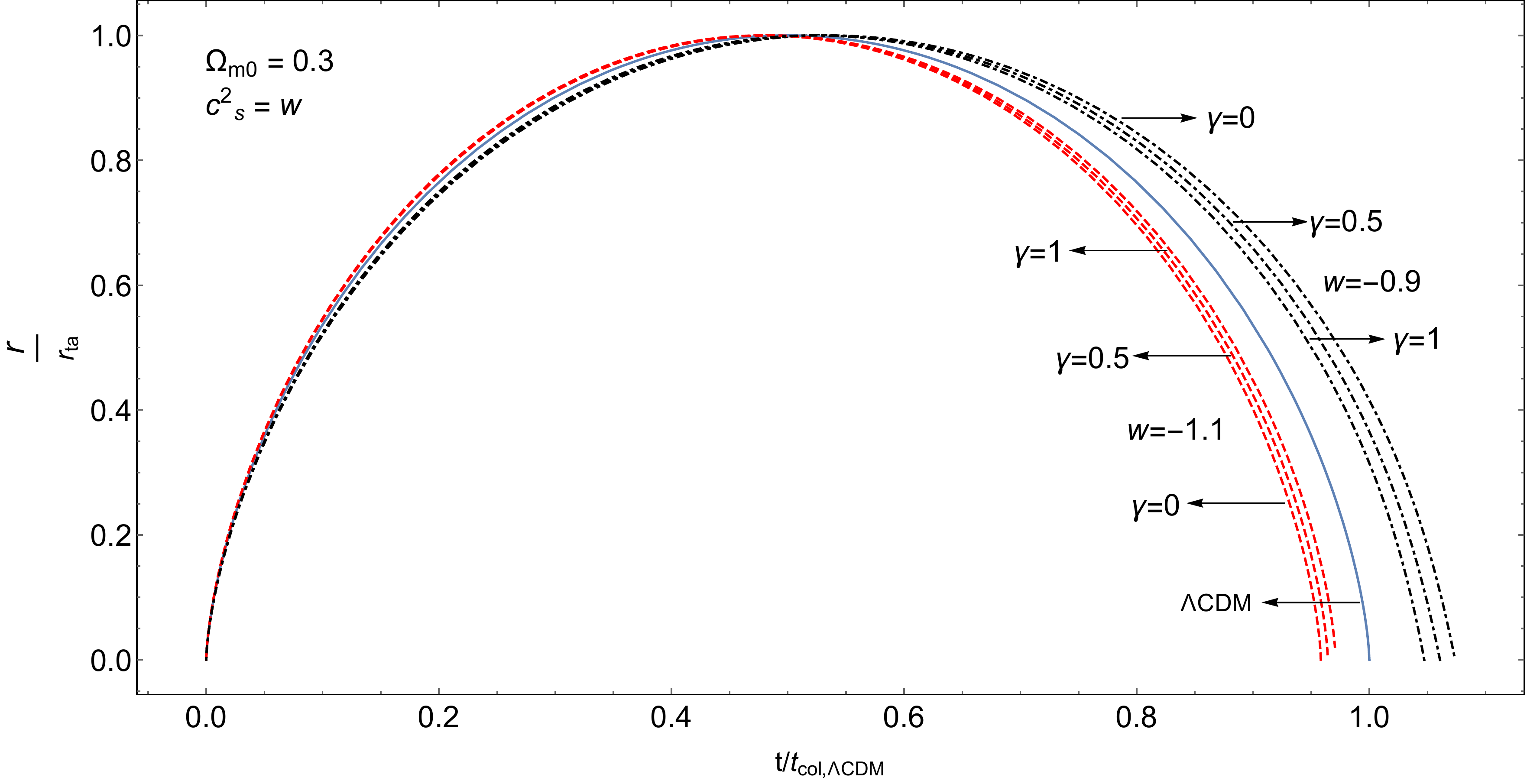}
\caption{Evolution of the scale radius of the collapsing sphere for  $c_s^2=w=\{ -0.9, -1.1\}$ and different values of $\gamma$. The solid blue line corresponds to the $\Lambda$CDM model.}
\label{radiia}
\end{figure}

\begin{figure}
\centering
\includegraphics[width=\textwidth]{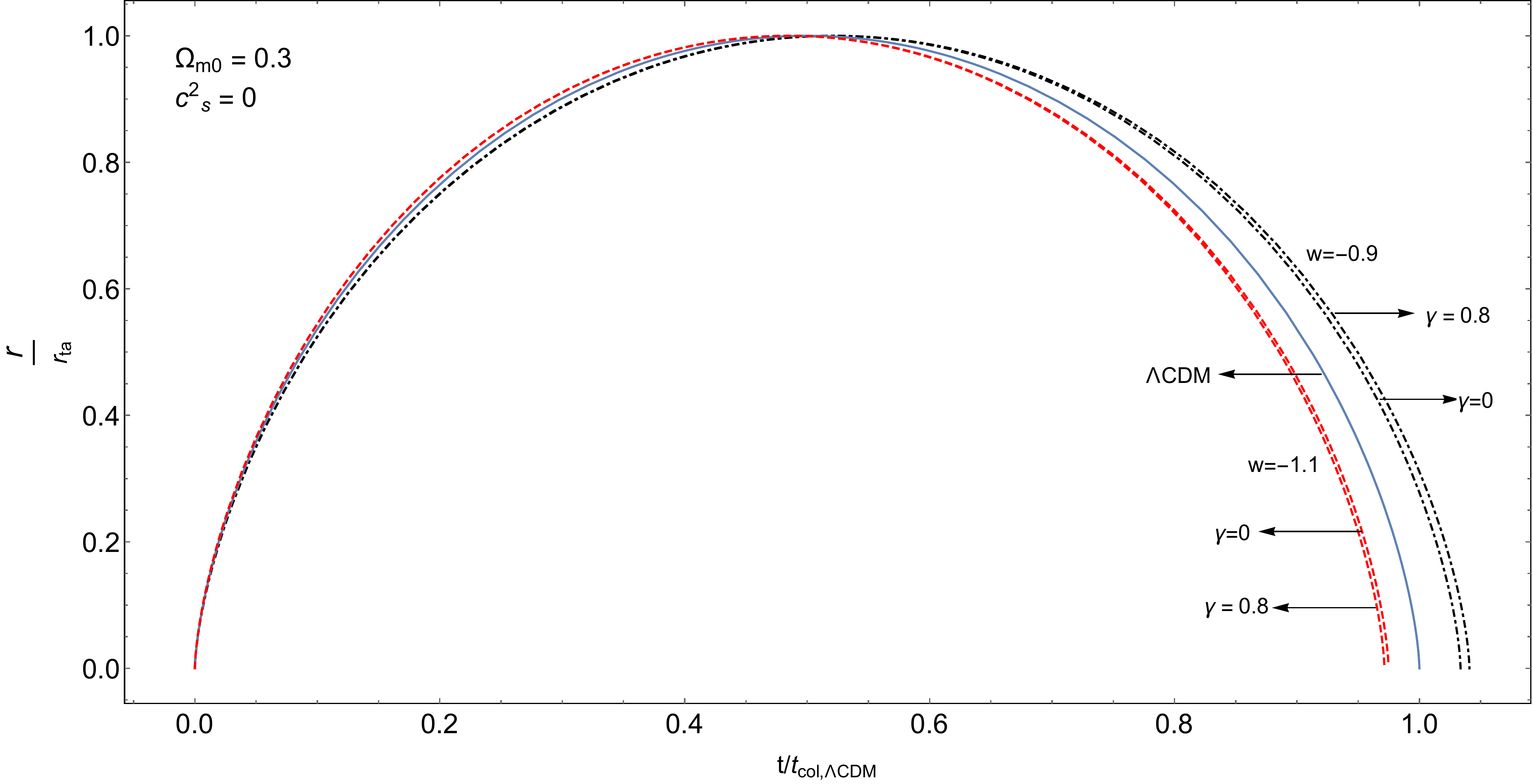}
\caption{Evolution of the scale radius of the collapsing sphere for  $c_s^2=0$, $w=\{ -0.9, -1.1\}$ and different values of $\gamma$. The solid blue line corresponds to the $\Lambda$CDM model.}
\label{radiib}
\end{figure}

\begin{figure}
\centering
\includegraphics[width=\textwidth]{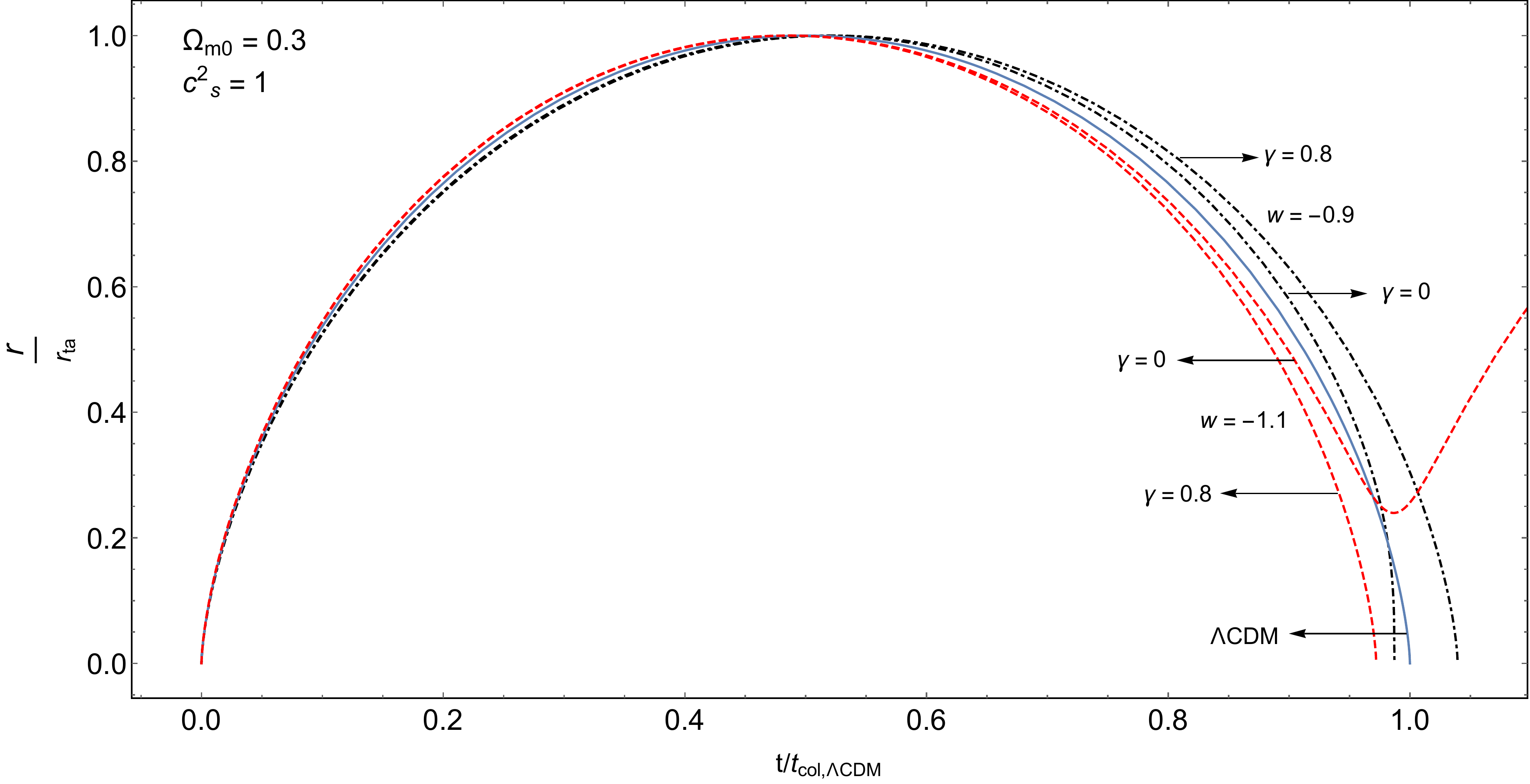}
\caption{Evolution of the scale radius of the collapsing sphere for  $c_s^2=1$, $w=\{ -0.9, -1.1\}$ and different values of $\gamma$. The solid blue line corresponds to the $\Lambda$CDM model. Note the non-collapsing curve ($\gamma=0, w=-1.1$). We also point out that the curve given by $\gamma=0$, $c_s^2=1$ and $w=-0.9$, that crosses $\Lambda$CDM close to the collapse, is also dissonant in Fig.~\ref{fig_deltacb}.}
\label{radiic}
\end{figure}

We now investigate the evolution of the spherical-region radius, as given by eq. (\ref{15a}). As mentioned before, the initial conditions for dark-matter perturbations in all models are fixed such that the collapse in $\Lambda$CDM  model always occurs at the present time. The initial conditions for dark-energy perturbations are given by Eq. (\ref{14a}). We point out some noteworthy features in a few particular cases:
\begin{itemize}
\item{$c_s^2 = w$ ($w^c= w$)\\}
The collapsing time $t_{col}$ is earlier than $\Lambda$CDM $t_{col,\Lambda CDM}$ only for phantom DE. This is a reasonable outcome, since $\delta_{de}<0$ if $w<-1$ (as mentioned above): the lack of DE in the clustered region accelerates the collapse.

For non-phantom, DE starts to dominate earlier when compared to $\Lambda$CDM  for any $\gamma$. On the other hand, a smaller $\gamma$ corresponds to a larger $\delta_{de}$, which will delay the collapse,  since in this case $\delta_{de}>0$.

There is no strong dependence on $\gamma$, except for a small drift towards $\Lambda$CDM when $\gamma \to 1$ (homogeneous DE), as expected. Besides, the term that inhibits  the collapse in Eq.~(\ref{4}), namely $\delta\rho_{de}+ 3\delta p_{de}$, although always present, will be less important in this limit. See Fig.~\ref{radiia}.

\item{$c_s^2 =0$\\}
We also get $t_{col}<t_{col,\Lambda CDM}$ only for phantom DE, as anticipated. The dependence on $\gamma$ is very weak. There is a slight drift {\it away} from $\Lambda$CDM as $\gamma\to 1$. Such opposite behavior (as compared to the previous case) happens because here $\delta p_{de}\equiv c_s^2 \delta\rho_{de} = 0$. Without any pressure support, the collapse is expedited if $\gamma\to 0$ and $w>-1$. Nevertheless, with phantom DE ($w<-1$), one has $\delta \rho_{de}<0$ and the clustering of DE (slightly) delays the collapse --- one can (barely) see the tiny shift to larger $t_{col}$ when $\gamma$ decreases from $0.8$ to $0$ in  Fig.~\ref{radiib}.

\item{$c_s^2=1$ (standard quintessence-like DE)\\}
As before, $t_{col}<t_{col,\Lambda CDM}$ for phantom DE but one can also expedite the collapse if $w>-1$. See  Fig.~\ref{radiic}. The most striking feature is the possibility to entirely prevent the collapse. This is not completely unexpected if there is enough stiff DE in the initial perturbation.The other ingredients for the bounce are phantom dark energy and no energy leaking. The full consequences of such behavior will be the subject of a future work. 

\end{itemize}

Here, the collapse time $t_{col}$ is defined as:
\begin{equation}{\label{tc}}
    t_{col}(w)= \int_{0}^{a_{c}}\frac{da}{H(w,a)a},
\end{equation}
where $a_{c}$ is the scale factor at collapse and $H(w,a)$ is the Hubble parameter of  the $w$CDM model. Of course, $t_{col}(w=-1)$ represents the collapse time of $\Lambda$CMD model, $t_{col,\Lambda CDM}$.
The curves $\gamma=1$ (homogeneous DE) from all the panels coincide, regardless of $c_s^2$, as expected.


\begin{figure}[h]
\centering
\includegraphics[width=\textwidth]{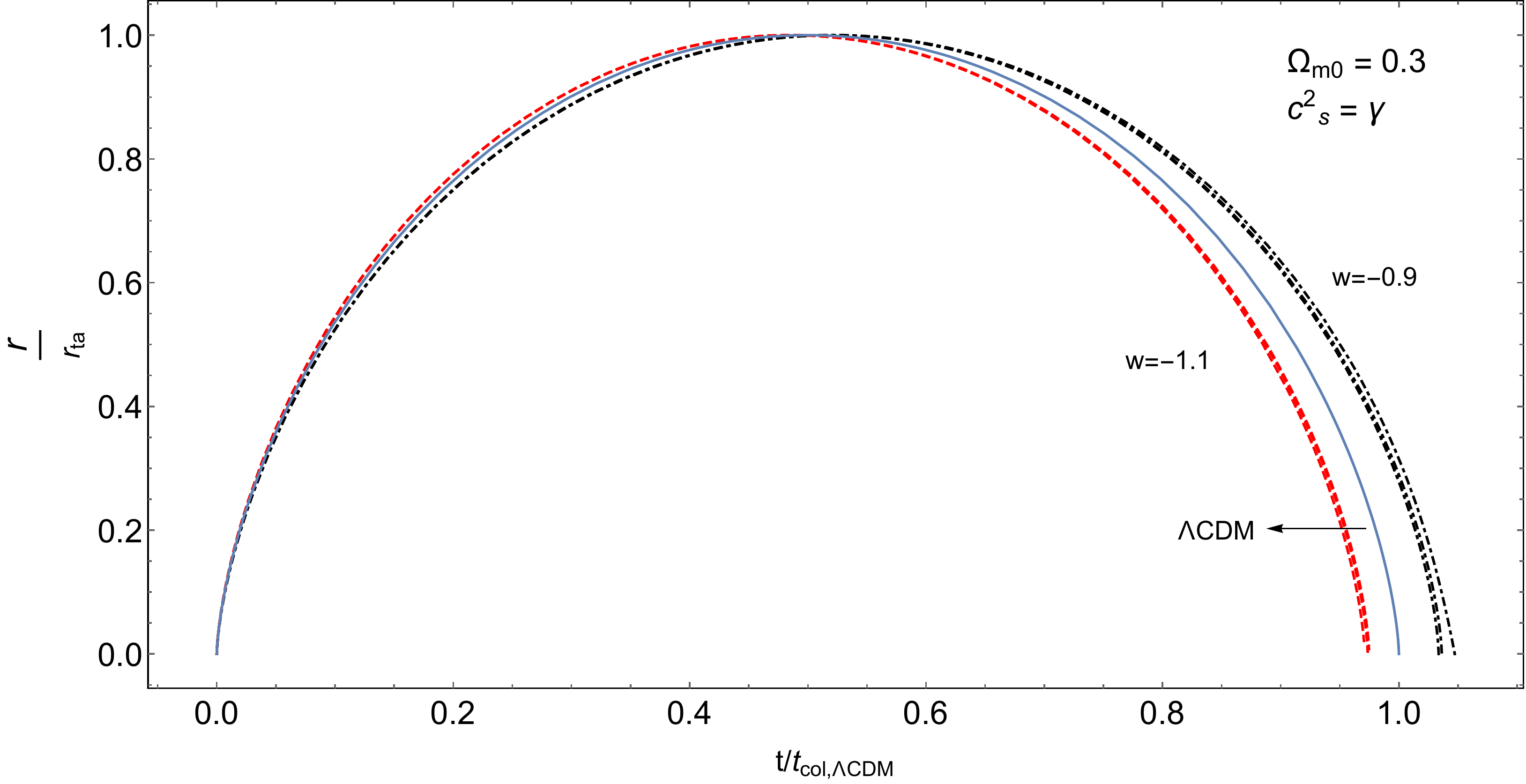}
\caption{Evolution of the scale radius of the collapsing sphere for $c_s^2=\gamma$, $w=\{ -0.9, -1.1\}$ and different values of $\gamma$ $=\{0, 0.8\}$. The solid blue line corresponds to the $\Lambda$CDM model. }
\label{radiid}
\end{figure}

\subsection{The critical contrast density}
\label{sec_critical}

\begin{figure}[h]
\centering
\includegraphics[width=\textwidth]{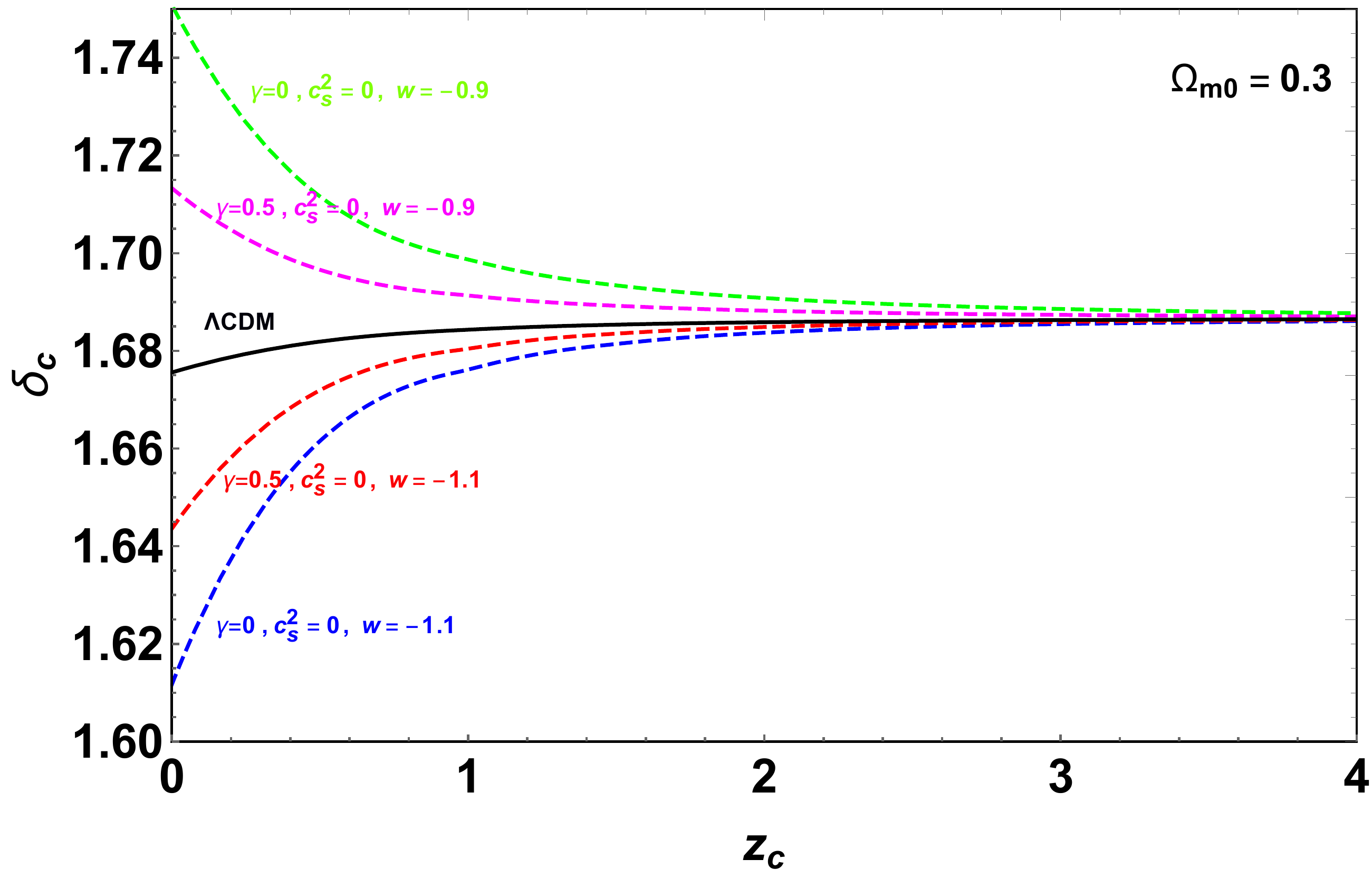}
\caption{Evolution of the critical contrast density for  $c^{2}_{s}=0$ and different values of $w$ and $\gamma$. The solid black line corresponds to $\Lambda$CDM model.}
\label{fig_deltaca}
\end{figure}

\begin{figure}
\centering
\includegraphics[width=\textwidth]{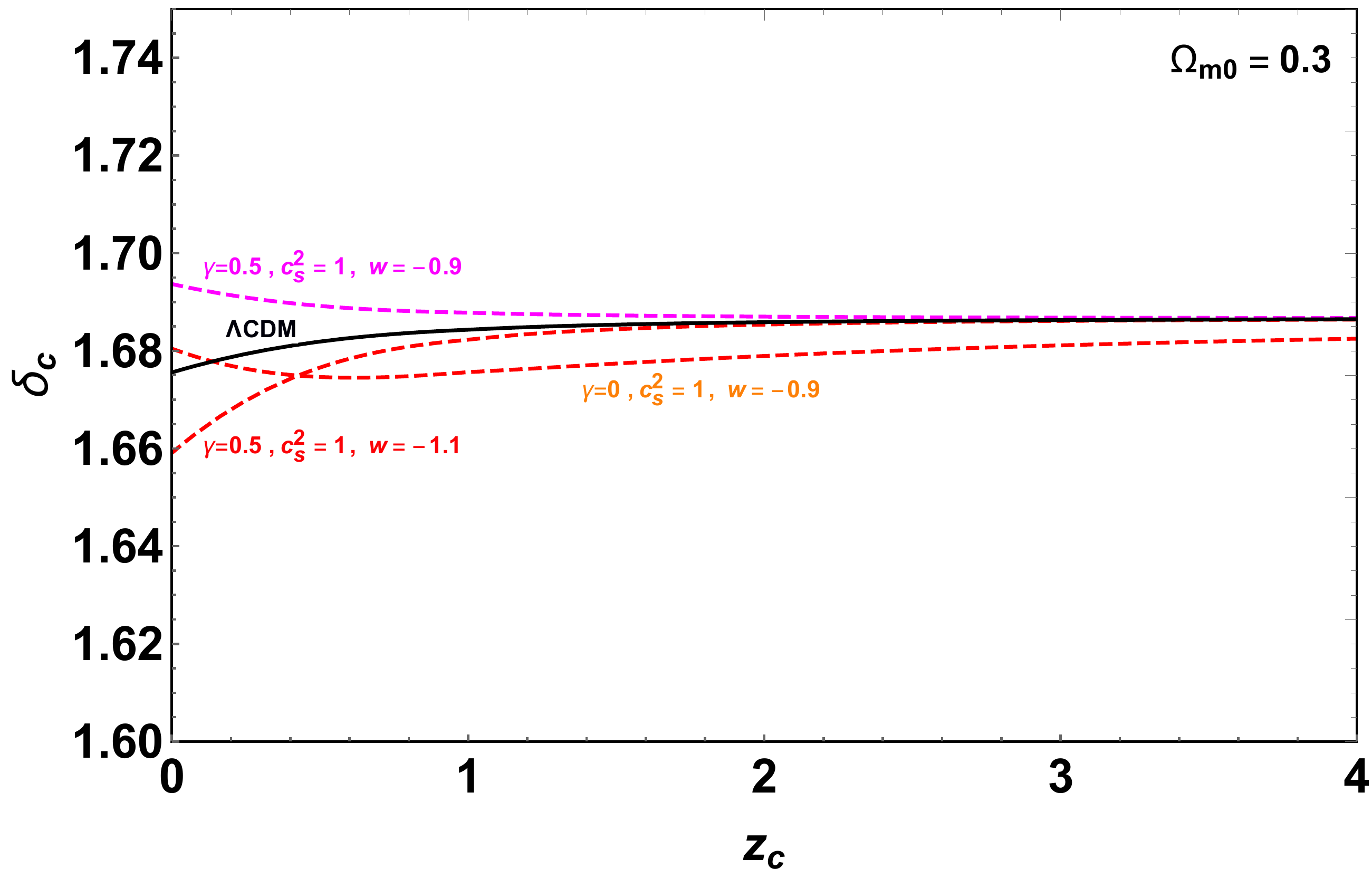}
\caption{Evolution of the critical contrast density for  $c^{2}_{s}=1$ and different values of $w$ and $\gamma$. The solid black line corresponds to $\Lambda$CDM model.}
\label{fig_deltacb}
\end{figure}

\begin{figure}
\centering
\includegraphics[width=\textwidth]{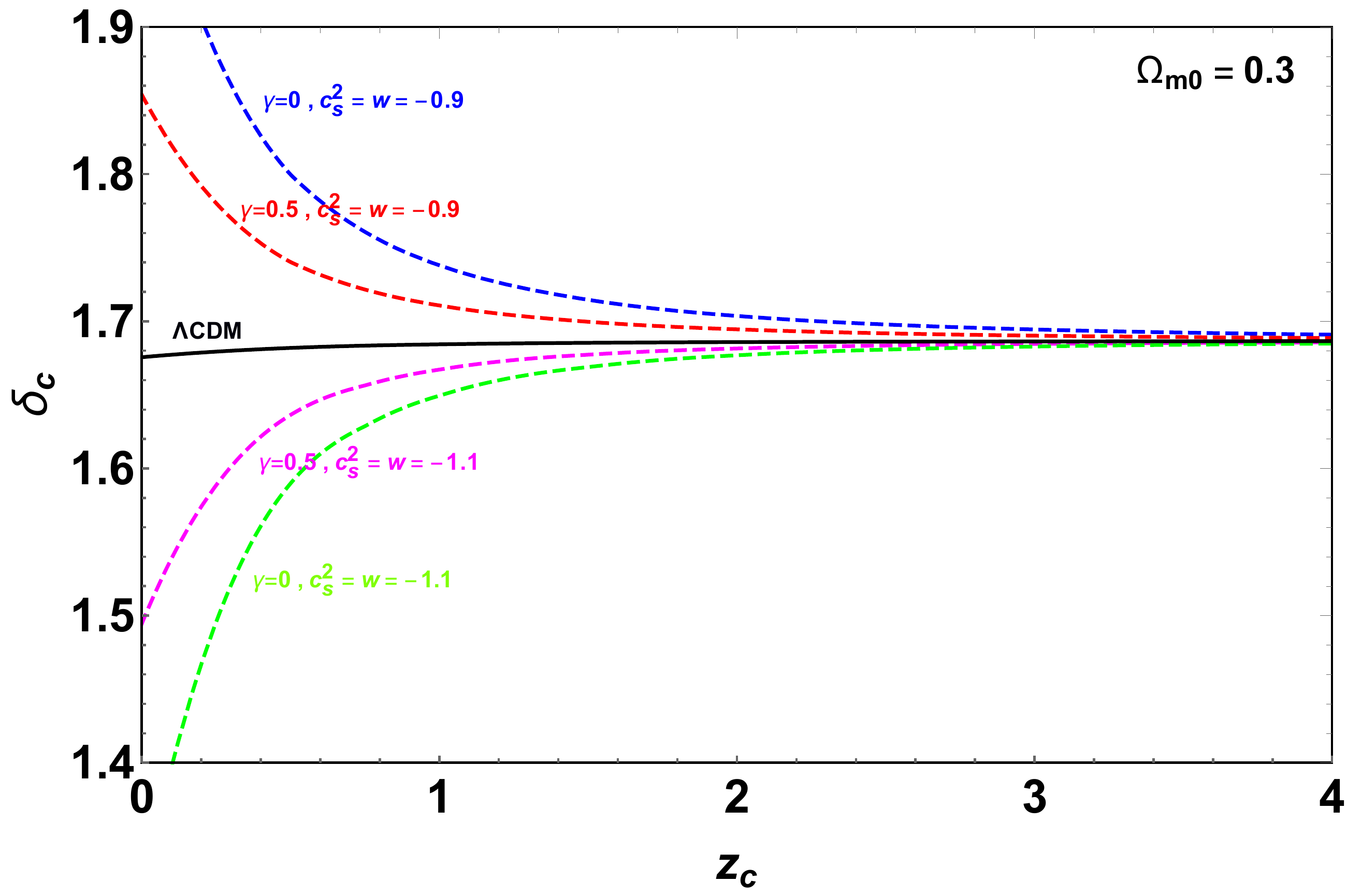}
\caption{Evolution of the critical contrast density for $c^{2}_{s}=w$ and different values of $w$ and $\gamma$. The solid black line corresponds to $\Lambda$CDM model in all panels. As before, here we find the largest deviations from $\Lambda$CDM.}
\label{fig_deltacc}
\end{figure}

\begin{figure}[h]
\centering
\includegraphics[width=\textwidth]{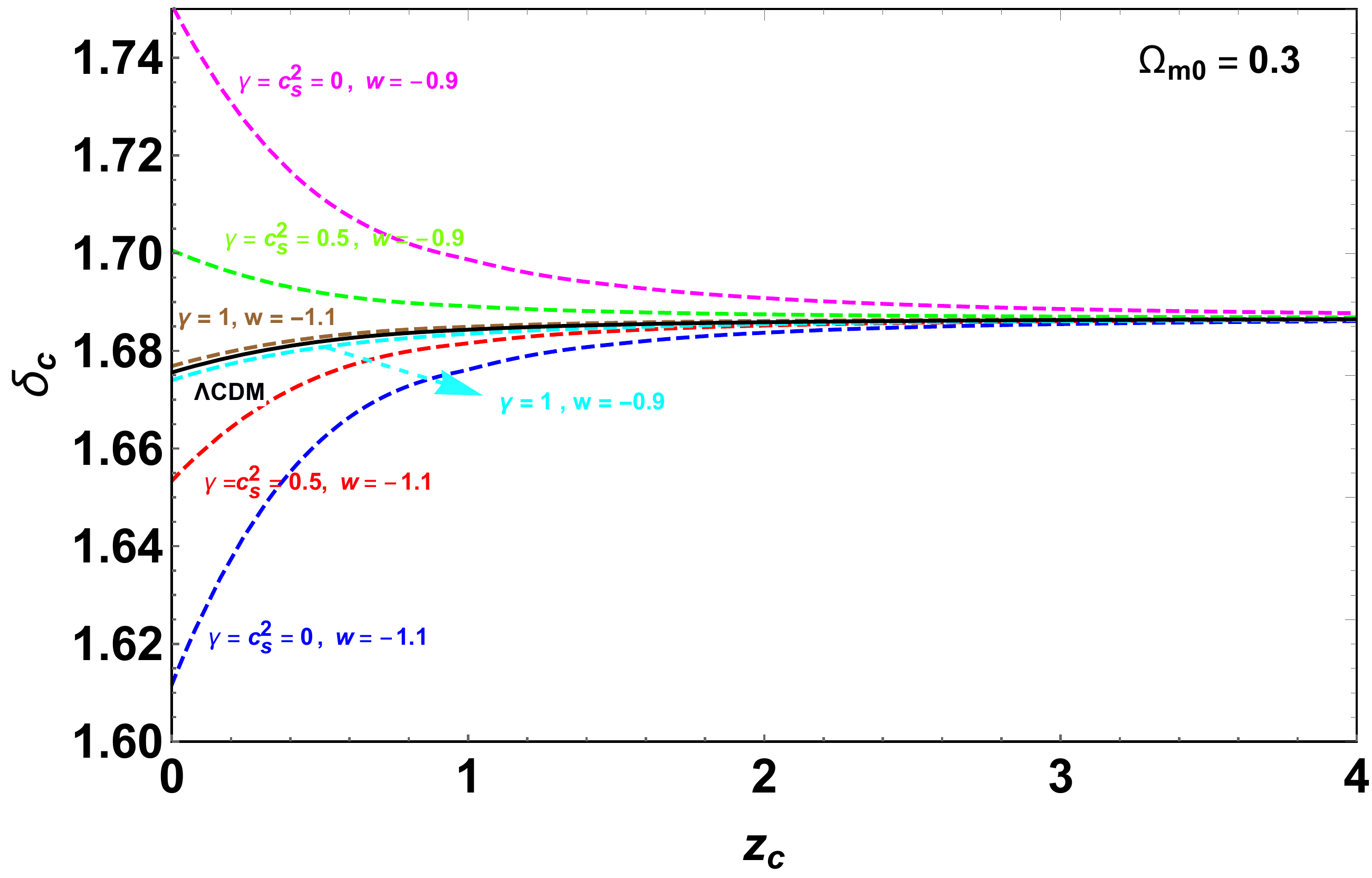}
\caption{Evolution of the critical contrast density for : $c^{2}_{s}=\gamma$ and different values of $w$ and $\gamma$. The solid black line corresponds to $\Lambda$CDM model in all panels.}
\label{fig_deltacd}
\end{figure}

As can be seen in Eq.~(\ref{4}) and from the discussions in the previous section, the  DE perturbations do contribute to the collapse. Therefore, the definition of the critical density contrast must be modified in order to take this contribution into account. So, let us consider the expression \cite{c6, 2011JCAP...03..047S}
\begin{equation}\label{17}
\delta_{tot}= \delta_{m}+ \frac{\Omega_{de}}{\Omega_{m}}\delta_{de},
\end{equation}
as the total perturbation. Note that, when $\Omega_{de} \rightarrow 0$, the conventional definition for the critical contrast is recovered. As usual, the critical contrast $\delta_c$ is determined by its linear evolution --- given by Eqs. (\ref{16a}) and (\ref{17a}) ---  at the collapse redshift $z_c$ (obtained from requiring that $r(z=z_c) \rightarrow 0 $):
\begin{equation}\label{18}
\delta_{c}=\delta_{tot}^{lin}(z_{c}).
\end{equation}
 Using the differential-radius method \cite{yo},  the dependence of $\delta_{c}$ with $z_c$ is shown in 
 Fig.~\ref{fig_deltaca}, \ref{fig_deltacb}, \ref{fig_deltacc}, and \ref{fig_deltacd} for different values of the free parameters $c_s^2$, $w$ and $\gamma$, and fixed $\Omega_{m0}=0.3$. 

Dark-energy overdensities ($\delta_{de}>0$) inhibits the growth of dark-matter perturbations ($\delta_m$) due to its repulsive nature. On the other hand, dark-energy underdensities ($\delta_{de}<0$) enhance the growth of $\delta_m$. The former case occurs in non-phantom models ($w>-1$), while the latter generally happens when $w<-1$. Indeed, as one can see in Figs.~\ref{fig_deltaca} to \ref{fig_deltacd}, the critical overdensity for a collapsing structure ($\delta_c$) is smaller in phantom cases. Therefore, one should expect an enhancement on the number of collapsed objects in this case. The choice $\gamma=0$ yields extreme variations of $\delta_c$ with respect to $\Lambda$CDM, because, in this case, there is no leakage of DE away from the collapsing regions, which maximizes its effects. 

In all the presented cases, $\delta_c$ tends to the expected EdS value at high $z_c$. Note also that $\delta_{c}$ is always larger (smaller) than the standard $\Lambda$CDM value for $w<-1$ ($w>-1$) and $\gamma\neq 1$ (i.e, in the presence of DE perturbations). When $\gamma=1$ (homogeneous DE), this behavior is inverted.  

The most striking feature in Fig.~\ref{fig_deltaca} ($c_s^2=0$) is the strong dependence of $\delta_c(z_c=0)$ on $w$ alone. That piece of information by itself reassures the importance of studying the critical density for breaking the degeneracy among different DE models. The dependence on $\gamma$ alone is not so strong ($\sim 2\%$). Changing both parameters at a time yields larger modifications on the curves, of course. 
The possibility of constraints on this parameters from observational data is beyond the scope of this paper. 

In Fig.~\ref{fig_deltacb}, where we keep $c_s^2=1$, we note once again the dependence on $w$, although about half as strong as in the previous case. One can notice a dissonant curve ($\gamma=0$, $c_s^2=1$, $w=-0.9$), which corresponds to the one that crosses over $\Lambda$CDM in Fig.~\ref{radiic}. It might be a sign of incompatibility of such parameters, since $\gamma=0$ means that there is no DE leaking away from the collapsing matter bubble, but at the same time $c_s^2=1$ corresponds to a stiff behavior of the former, which should (at least) delay the DE collapsing process. 

 The strongest dependence of $\delta_c(z_c=0)$ on the parameters is observed in Fig.~\ref{fig_deltacc}, where we keep $c_s^2=w$. Observe also that, for  
 larger $w$,  $\delta_c(z_c=0)$ rapidly increases. 
 For (non)phantom DE, a (larger) smaller $\gamma$ decreases $\delta_c(z_c=0)$.
  On the other hand, a larger failure on energy conservation (i.e, larger $\gamma$) in the collapsing region does move any of the curves towards $\Lambda$CDM. 
  
  The cases $c_s^2=\gamma$ are depicted in Fig.~\ref{fig_deltacd}. As expected, the curves tend to $\Lambda$CDM whenever $\gamma\to 1$, regardless of the values of $c_s^2$.
 
 We also notice that if $\gamma= 1$ (without DE perturbation), the phantom-DE curve is slightly above $\Lambda$CDM, as opposed to all the other cases presented here. The non-phantom is also inverted (below $\Lambda$CDM in this case alone).


\section{Conclusions}
\label{sec_conclusions}

In summary, we have shown the non-linear equations that describe the evolution of the perturbations for both the dark matter and dark energy in the SC model when the clustering fraction of the latter is defined by a parameter $\gamma$, which consequently also models the lack of energy conservation in the collapsing region.

We have determined the  critical contrast density $\delta_c$ for different values of $\gamma$, obtaining larger values for stronger DE clustering. The largest discrepancies from $\Lambda$CDM happen when $c_s^2=w$ (both clustered and smooth DE have the same EoS) and $\gamma=0$ (fully clustered DE).

In a next paper, we will explore the consequences of the results presented here, namely deviations on the number density of collapsed objects, and the possibility of constraining the free parameters with current and future observational data.

\section*{ACKNOWLEDGMENTS}
D.H. acknowledges financial support from CAPES.

\bibliographystyle{unsrt}
\bibliography{biblio1}

\end{document}